\documentclass[prb,twocolumn,amsmath,amssymb,superscriptaddress,showpacs]{revtex4-1}

\usepackage{bm}
\usepackage{graphicx}
\usepackage{times}
\usepackage[colorlinks,citecolor=blue,linkcolor=blue]{hyperref}
\newcommand{\eps}{\varepsilon}

\begin{document}

\title{Strain tuning of topological band order in cubic semiconductors}

\author{Wanxiang Feng}

\affiliation{Materials Science \& Technology Division, Oak Ridge National Laboratory,
Oak Ridge, Tennessee 37831, USA}
\affiliation{Department of Physics and Astronomy, University of Tennessee, Knoxville,
Tennessee 37996, USA}
\affiliation{Beijing National Laboratory for Condensed Matter Physics and Institute
of Physics, Chinese Academy of Sciences, Beijing 100190, China}

\author{Wenguang Zhu}

\affiliation{Department of Physics and Astronomy, University of Tennessee, Knoxville,
Tennessee 37996, USA}
\affiliation{Materials Science \& Technology Division, Oak Ridge National Laboratory,
Oak Ridge, Tennessee 37831, USA}

\author{Hanno H. Weitering}

\affiliation{Department of Physics and Astronomy, University of Tennessee, Knoxville,
Tennessee 37996, USA}
\affiliation{Materials Science \& Technology Division, Oak Ridge National Laboratory,
Oak Ridge, Tennessee 37831, USA}

\author{G. Malcolm Stocks}

\affiliation{Materials Science \& Technology Division, Oak Ridge National Laboratory,
Oak Ridge, Tennessee 37831, USA}

\author{Yugui Yao}

\affiliation{School of Physics, Beijing Institute of Technology, Beijing 100081, China}
\affiliation{Beijing National Laboratory for Condensed Matter Physics and Institute
of Physics, Chinese Academy of Sciences, Beijing 100190, China}

\author{Di Xiao}
\affiliation{Materials Science \& Technology Division, Oak Ridge
  National Laboratory, Oak Ridge, Tennessee 37831, USA}

\begin{abstract}
We theoretically explore the possibility of tuning the topological order of cubic diamond/zinc-blende semiconductors with external strain. Based on the tight-binding model, we analyze the evolution of the cubic semiconductor band structure under hydrostatic or biaxial lattice expansion, by which a generic guiding principle is established that lattice \emph{expansion} can induce a topological phase transition of small band-gap cubic semiconductors via a band inversion, and further breaking of the cubic symmetry leads to a topological insulating phase. Using density functional theory calculations, we demonstrate that a prototype topological trivial semiconductor, InSb, is converted to a nontrivial topological semiconductor with a $2\%\--3\%$ biaxial lattice expansion.
\end{abstract}


\maketitle

\section{Introduction}

The recent discovery of topological insulators has generated great interest in the fields of condensed matter physics and materials science, largely driven by their exotic surface electronic properties.~\cite{Qi2010,Qi2011,Hasan2010}  These materials are insulating in the bulk, but support topologically protected helical edge or surface state in two- or three-dimensional systems.~\cite{Kane2005a,Kane2005b,Qi2010,Qi2011,Hasan2010}  The surface states are predicted to have special properties that could be useful for practical applications in spintronics and quantum computation.~\cite{Moore2010}  To date, topological states have been observed in a number of materials,~\cite{Teo2008,Hsieh2008,Bernevig2006,Konig2007,Brune2011,Fu2007a,Zhang2009,Xia2009,Chen2009,Xiao2010a,Chadov2010,Lin2010a,Feng2011,Feng2010,Al-Sawai2010,Liu2011a,Gofryk2011,Lin2010b,Yan2010a,Chen2010,Sato2010,Kuroda2010,Sun2010,Yan2010b,Kim2010,Jin2011,Zhang2011,Chen2011,Liu2011} opening the door for detailed and systematic investigations of various topological phenomena in laboratories.  While the essential features of topological insulators have been established, our understanding of the fundamental physics in relation to basic material properties is far from complete.  This type of knowledge will be crucial in potential device applications.

In this work, we theoretically demonstrate the possibility of tuning the topological order in cubic semiconductors by applying external strain, using InSb as a prototype system.  Our choice of the material system is motivated by the fact that many technologically important semiconductors crystalize in one of the cubic crystal structures, including the group-IV elements in the diamond structure, and the III-V and II-VI compounds in the zinc-blende structure.  Ternary compound semiconductors with the tetrahedral coordination, such as half-Heuslers~\cite{Xiao2010a,Chadov2010,Lin2010a} and chalcopyrites~\cite{Feng2011}, also belong to this category.  We first establish, based on tight-binding analysis, that lattice \emph{expansion} can induce a topological phase transition of small band-gap cubic semiconductors via a band inversion, and further breaking of the cubic symmetry leads to a topological insulating phase. Using density functional theory calculations, we then demonstrate that InSb enters into a topologically nontrivial state at a reasonable biaxial lattice expansion around $2\%\--3\%$.  The relationship between mechanical strain and topological order has been discussed in layered bismuth compounds~\cite{Young2011} as well as half-Heuslers~\cite{Xiao2010a,Chadov2010,Lin2010a}.  Our analysis provide a physical picture of these findings.

\section{General Considerations}

The band topology of materials with time-reversal symmetry is characterized by the so-called $Z_2$ topological index.  In general, the calculation of the $Z_2$ index requires the knowledge of the occupied Bloch wave functions in the entire Brillouin zone because it is a global property of the energy bands.  However, in most practical situations the $Z_2$ index can be determined by counting the number of band inversions at the time-reversal invariant $\bm k$-points.  In passing from a topologically trivial to nontrivial state, a band inversion, i.e., switching of occupied and unoccupied bands with opposite parity around the Fermi level, must happen.  If an odd number of band inversions occur within the Brillouin zone, then the material could be a topological insulator.

\begin{figure}[b]
\includegraphics[width=\columnwidth]{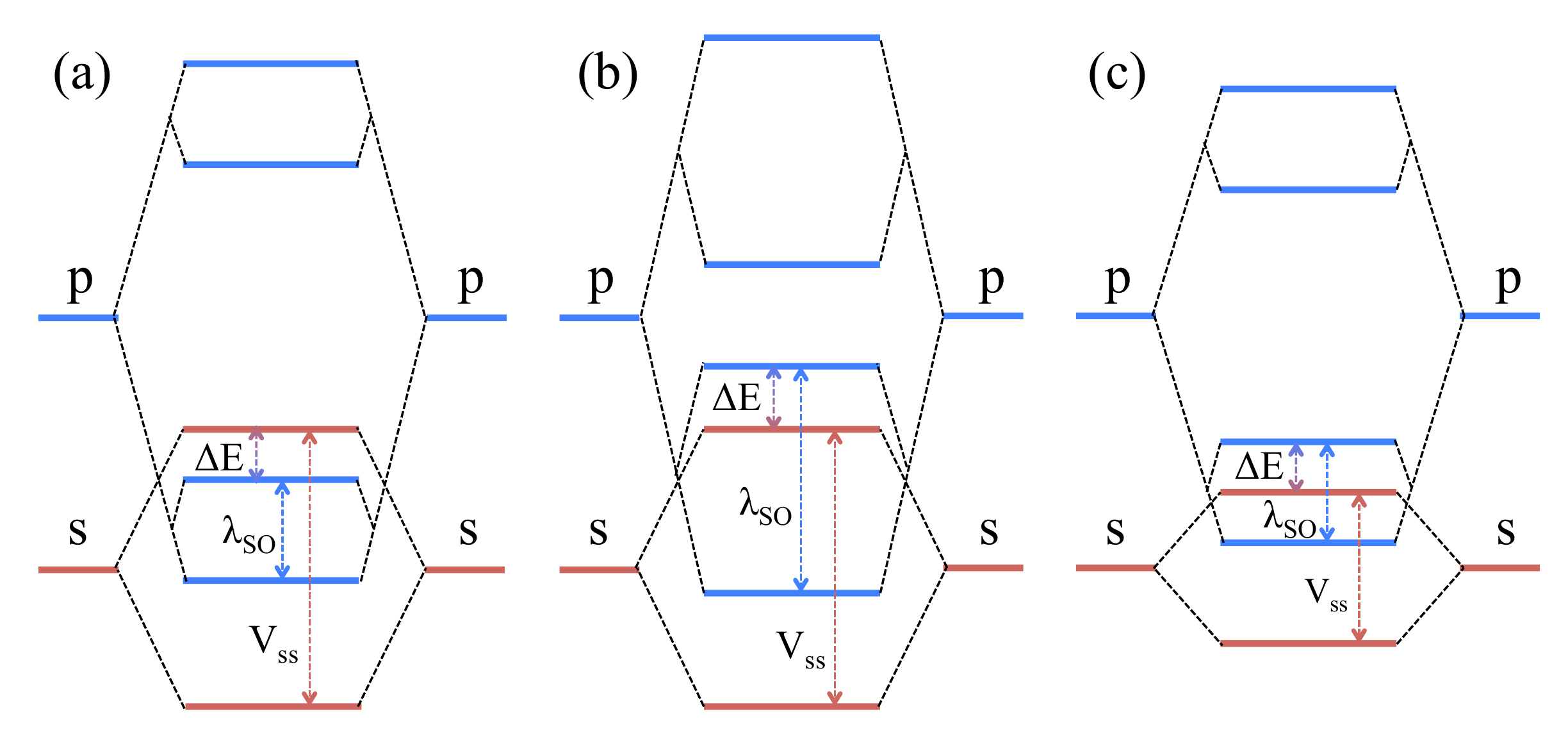}
\caption{\label{fig:order}(Color online) The band order of cubic diamond/zinc-blende semiconductors at the $\Gamma$ point with $\lambda_\text{SO} > 0$.  (a) The normal band order.  (b)-(c) The inverted band order can be obtained from the normal band order by either increasing the spin-orbit coupling $\lambda_\text{SO}$ (b) or by decreasing the coupling potentials $V_{ss}$ and $V_{pp}$ (c).  The band inversion strength is defined as $\Delta E=E_{\Gamma_6}-E_{\Gamma_8}$, which has positive value in normal band order and negative value in inverted band order.}
\end{figure}

We now specialize to zinc-blende semiconductors.  In a typical band structure, the low-energy electronic properties is dominated by bands around the $\Gamma$ point.  Away from the $\Gamma$ point the valence and conduction bands are well separated.  Therefore we only need consider the band order at $\Gamma$.  The relevant states are the $\Gamma_6$ anti-bonding state of the $s$-orbitals, and the $\Gamma_7$ and $\Gamma_8$ bonding states of the $p$-orbitals.  Using the tight-binding model,~\cite{Slater1954,Chadi1977} the band energies of these states are obtained
\begin{align}
E_{\Gamma_6} &= \frac{\eps_{s1} + \eps_{s2}}{2}
+ \sqrt{\Bigl(\frac{\eps_{s1} - \eps_{s2}}{2}\Bigr)^2 + V_{ss}^2} \;, \\
E_{\Gamma_7} &= \frac{\eps_{p1} + \eps_{p2}}{2}
- \sqrt{\Bigl(\frac{\eps_{p1} - \eps_{p2}}{2}\Bigr)^2 + V_{pp}^2}
- \lambda_\text{SO} \;, \\
E_{\Gamma_8} &= \frac{\eps_{p1} + \eps_{p2}}{2}
- \sqrt{\Bigl(\frac{\eps_{p1} - \eps_{p2}}{2}\Bigr)^2 + V_{pp}^2}
+ \frac{1}{2}\lambda_\text{SO} \;,
\end{align}
where $\eps_{s1,2}$ and $\eps_{p1,2}$ are the $s$- and $p$-orbital energies of the cation and anion, respectively, $V_{ss}$ and $V_{pp}$ are the coupling potentials of the $s$-$s$ and $p$-$p$ bonds, and $\lambda_\text{SO}$ is the spin-orbit coupling strength.  Note that $\Gamma_6$ and $\Gamma_{7,8}$ states have opposite parity.

Usually, the $\Gamma_6$ state is located above the $\Gamma_{7,8}$ states, forming the conduction band minimum.  This is the so-called normal band order [see Fig.~\ref{fig:order}(a)].    There are two ways to change the band order.  (i) If the spin-orbit coupling strength is increased, depending on the sign of $\lambda_\text{SO}$, either the $\Gamma_8$ ($\lambda_\text{SO} > 0$) or the $\Gamma_7$ ($\lambda_\text{SO} < 0$) state could rise above the $\Gamma_6$ state, realizing the necessary band inversion [see Fig.~\ref{fig:order}(b)].  (ii) In addition to varying $\lambda_\text{SO}$, which is an intrinsic property of the material, the band order can be also changed by varying the coupling potentials $V_{ss}$ and $V_{pp}$ [see Fig.~\ref{fig:order}(c)].  Increasing the lattice constant leads to a decrease of the coupling potentials, which lowers the $\Gamma_6$ anti-bonding state and raises the $\Gamma_{7,8}$ bonding states.  At sufficiently large lattice constant, the band inversion can be realized.  Thus in the design of topological insulators, both factors should be taken into account.

The relation between the band gap and lattice constant has been discussed previously in the context of band gap pressure coefficient.~\cite{Wei1999,Li2006}  In a more realistic situation, there are other factors that could affect the band positions, such as the $p$-$d$ coupling in the compounds with active $d$-orbitals in the valence bands, as well as the kinetic energy effect under pressure.  A detailed analysis can be found in Refs. \onlinecite{Wei1999,Li2006}; the general trend we discussed above still holds.

\begin{figure}
\includegraphics[width=\columnwidth]{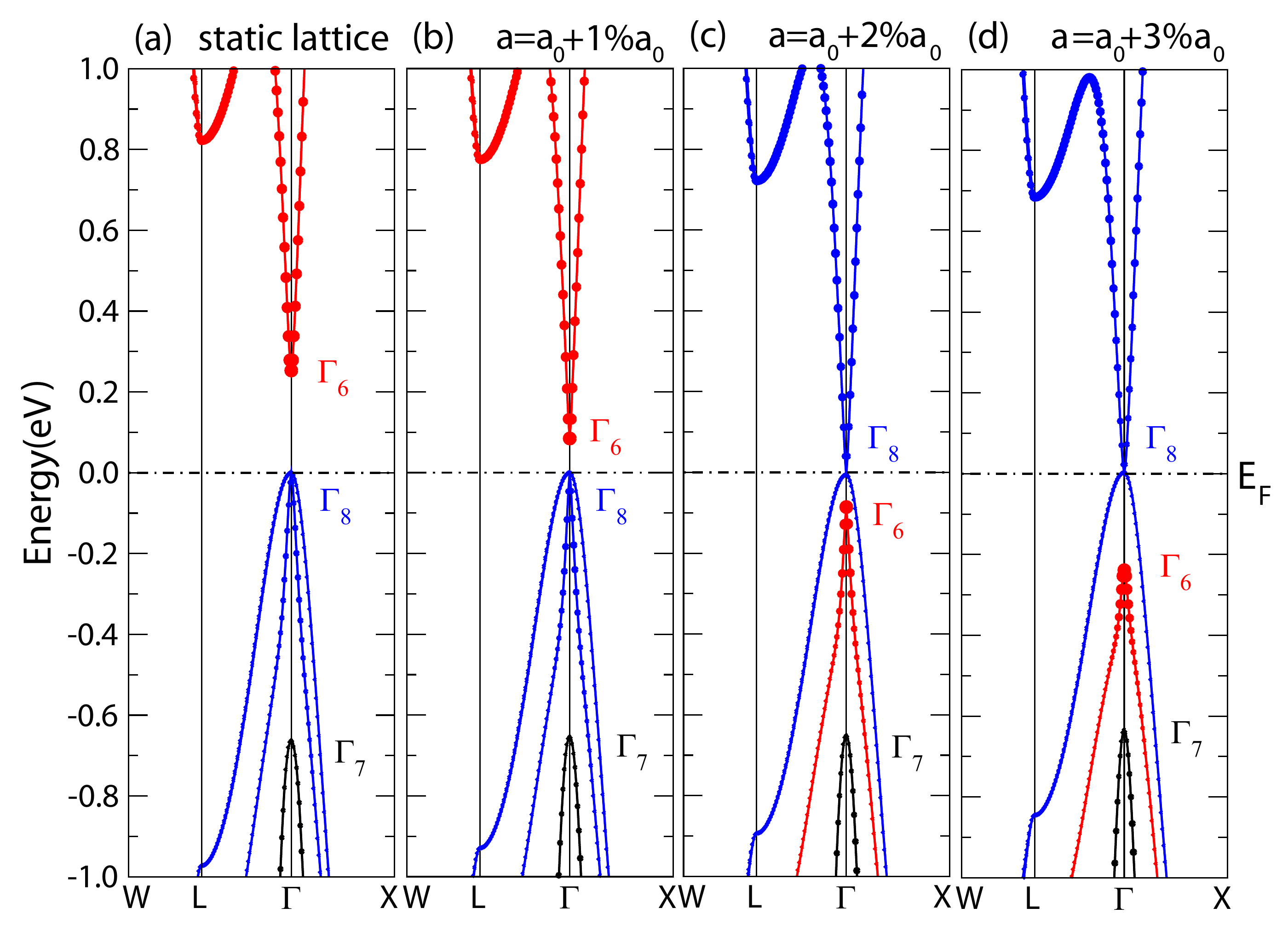}
\caption{(Color online) The band structures of InSb at the equilibrium lattice constant (a) and with 1\% (b), 2\% (c), and 3\% (d) hydrostatic lattice expansions. The $\Gamma_{6}$, $\Gamma_{7}$, and $\Gamma_{8}$ state are denoted by red, black, and blue color, respectively. The size of dots is proportional to the probability of $s$-orbit projection.}
\label{fig:hydrostatic}
\end{figure}

\begin{figure}
\includegraphics[width=\columnwidth]{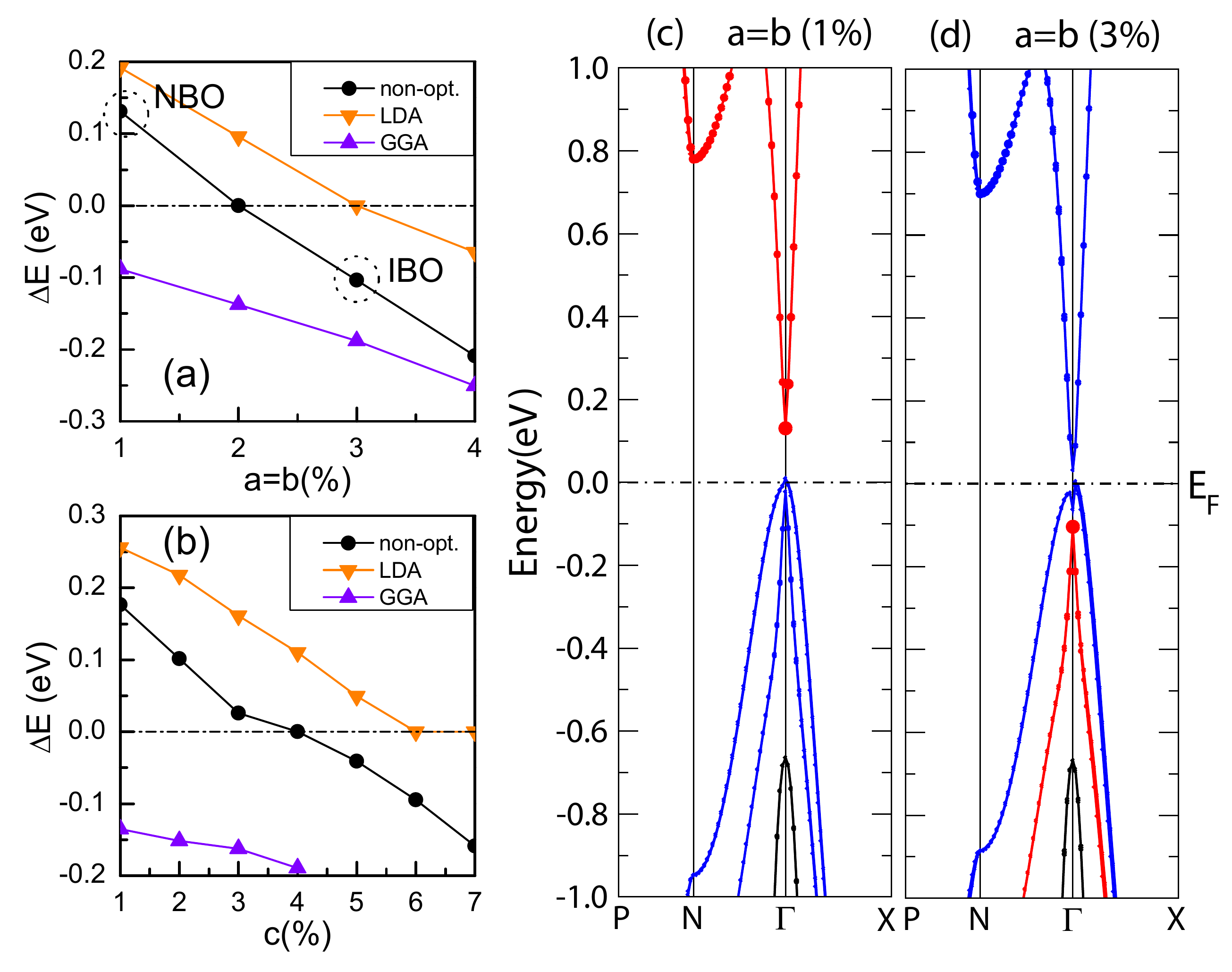}
\caption{(Color online) The evolution of band order of InSb with nonhydrostatic expansion (a) in $ab$-plane and (b) along $c$-axis. The black solid circles are the results without lattice optimization, orange down-diamonds and purple up-diamonds are optimized by LDA and GGA potential receptively. (c) and (d) are normal band order and inverted band order, which correspond to the states marked by dotted circles in (a). The size of dots is proportional to the probability of $s$-orbit projection.}
\label{fig:uniaxial}
\end{figure}

Finally, if $E_{\Gamma_8} > E_{\Gamma_6}$ in the inverted band structure ($\lambda_\text{SO} > 0$), the twofold degenerate ${\Gamma}_6$ state is fully occupied while the fourfold degenerate ${\Gamma}_8$ state is half filled, resulting in a zero band-gap semiconductor [for example, see Fig.~\ref{fig:hydrostatic}(c) and \ref{fig:hydrostatic}(d)]. To realize a truly insulating ground state, a possible way is to apply a biaxial strain so as to break the cubic symmetry and thereby open a band gap at $\Gamma$ point  [see Fig.~\ref{fig:uniaxial}(d)].

\begin{table}
\caption{The relative energy changes of states $E_{\Gamma_6}$, $E_{\Gamma_7}$, and $E_{\Gamma_8}$ under the hydrostatic lattice expansions ($a=b=c$) ranging from $0\%$ to $4\%$.}\label{tab:energy1}
\begin{ruledtabular}
\begin{tabular}{cccc}
\multicolumn{1}{c}{\textrm{a(\%)}} &
\multicolumn{1}{c}{\textrm{V(Bohr$^3$)}} &
\multicolumn{1}{c}{\textrm{$E_{\Gamma_6}-E_{\Gamma_8}$(eV)}} &
\multicolumn{1}{c}{\textrm{$E_{\Gamma_8}-E_{\Gamma_7}$(eV)}} \\
\hline

 0  &  458.92429  &  0.25321  &  0.66261  \\
 1  &  472.83016  &  0.08445  &  0.65357  \\
 2  &  487.01414  & -0.08042  &  0.64499  \\
 3  &  501.47899  & -0.24181  &  0.63672  \\
 4  &  516.22745  & -0.39837  &  0.62886  \\

\end{tabular}
\end{ruledtabular}
\end{table}

\section{Methodology}

Having established the general tuning principle, next we verify whether the anticipated topological phase transition can happen in real materials at a reasonable lattice expansion. We choose InSb as a prototype system because it has a relatively small band gap of 0.235 eV,\cite{Madelung2004} indicating that InSb is on the verge of becoming a topological insulator.

The band structure calculations in this work were performed using full-potential linearized augmented plane-wave method\cite{Singh1994,Blugel2006} implemented in package \textsc{wien2k}.\cite{Blaha2001}  A converged ground state was obtained using $10^4$ $\bm k$-points in the first Brillouin zone with $R_\text{MT} K_\text{max} = 9.0$, where $R_\text{MT}$ represents the smallest muffin-tin radius and $K_\text{max}$ is the maximum size of reciprocal-lattice vectors. The muffin-tin radius was set to $2.5$ Bohr for both  In and Sb atoms. Wave functions and potentials inside the atomic sphere are expanded in spherical harmonics up to $l=10$ and $4$, respectively.  Spin-orbit coupling is included by a second-variational procedure,~\cite{Singh1994} where states up to 9 Ry above Fermi level are included in the basis expansion.  The modified Becke-Johnson exchange potential together with local-density approximation for the correlation potential (MBJLDA)~\cite{Tran2009} was used to obtain the band structures.  The MBJDA potential can effectively mimic the behavior of orbital-dependent potential around the band gap, and it is expected to obtain accurate positions of states near the band edge~\cite{Tran2009}.  In fact,  both standard  local-density approximation (LDA)\cite{Kohn1965,Perdew1992} and generalized gradient approximation (GGA)\cite{Perdew1996} calculations predict that InSb is a zero-band gap semiconductor with an inverted band order, while the MBJLDA calculation produce a band gap about 0.253 eV, which is in excellent agreement with the experimental value of 0.235 eV.~\cite{Madelung2004}  The MBJLAD method has been applied for small band-gap half-Heuslers in the context of topological insulators~\cite{Feng2010,Al-Sawai2010} and yields favorable comparison with more sophisticated GW method.~\cite{Vidal2011}

We consider two types of lattice expansion. The first is the hydrostatic lattice expansion,  equally increasing the lattice constants along all the three axes. This situation is more of academic interest rather than for practical applications. In the second case, we consider the biaxial lattice expansion, where the crystal structure is expanded in the $ab$-plane and relaxed in the $c$-axis. This can be achieved by growing the sample on a substrate with a larger lattice constant. Although the MBJLDA potential gives very accurate band gap and band order, it has no corresponding definition about its energy functional.~\cite{Tran2009}  Hence, both the LDA and GGA for the exchange-correlation potential were used to optimize the lattice constants.

\section{Result and Discussion}

The crystal structure of InSb is zinc-blende with space group $F\bar{4}3m$ (No. 216).
The unit cell of InSb contains two inequivalent atoms, In and Sb, with the fractional coordinates (0,0,0) and (0.25,0.25,0.25), respectively. The initial equilibrium lattice constant of 6.479 $\text{\AA}$ is taken from previous  experimental measurements.\cite{Madelung2004}

Figure~\ref{fig:hydrostatic} shows the band structures of InSb at its equilibrium lattice constant and under hydrostatic lattice expansion ranging from $1\%$ to $3\%$.  The relative energy changes of the states $E_{\Gamma_6}$, $E_{\Gamma_7}$, and $E_{\Gamma_8}$ are listed in Table~\ref{tab:energy1}.  We find that, as anticipated, the band inversion takes place upon lattice expansion, somewhere between 1\% and 2\%.  In addition, we find that the spin-orbit coupling strength, characterized by $E_{\Gamma_8} - E_{\Gamma_7}$, stays almost a constant with a slight decrease as the lattice constant increases.  Hence the evolution of the band order during lattice expansion is dominated by the change of the coupling potentials $V_{ss}$ and $V_{pp}$.

Although a hydrostatic expansion already converts the normal band order into the inverted band order, a nonhydrostatic strain is still needed to creat a band gap at $\Gamma$ point by breaking the cubic symmetry.  As shown in Fig.~\ref{fig:uniaxial}, we investigate the evolution of band order with the nonhydrostatic expansion both in $ab$-plane and along the $c$-axis.  Note that in the tetragonal symmetry the band inversion strength $\Delta E$ is redefined as the energy difference between the $s$-like states and the top of valence bands.  First, we consider the lattice expansion in the $ab$-plane with fixed lattice constant along the $c$-axis (without lattice optimization).  The $\Delta E$ change from positive values to negative values as shown in Fig. \ref{fig:uniaxial}(a) (black solid circles). The band structures of two typical states are presented in Fig.~\ref{fig:uniaxial}(c) and \ref{fig:uniaxial}(d).  One is normal band order with $1\%$ expansion [Fig.~\ref{fig:uniaxial}(c)] and another is inverted band order with $3\%$ expansion [Fig.~\ref{fig:uniaxial}(d)].  Then, we consider the lattice optimization along the $c$-axis by using both LDA and GGA potential.  It is well known that LDA usually underestimates the lattice constant and GGA overestimates it, hence these results give us a lower and upper bound.  The true transition point for the expansion in $ab$-plane is in the range of $2\%\sim3\%$. Finally, if we expand the lattice along $c$-axis, there exist similar topological order transition. The only quantitative difference is that a larger lattice expansion is needed, as shown in \ref{fig:uniaxial}(b).

%The biaxial expansion in the $ab$-plane is more relevant in a realistic experimental setup. But a nature question is that how large the tensile biaxial stress is needed to satisfy the critical lattice expansion for the topological phase transition? This can be simply estimated through the variational relationship between the total energy and the volume from density functional theory calculations, i.e., $P=-dE/dV$. For the upper limit of the predicted critical biaxial lattice expansion for InSb ($3\%$), the corresponding pressure is estimated to be in the range of $-0.9$ GPa from LDA calculations and $-0.1$ GPa from GGA calculations, respectively. This magnitude of pressure can be easily achieved experimentally. In addition,

Finally, we mention the possibility that the tensile biaxial expansion may also be realized by growing the cubic semiconductor on the top of a piezoelectric substrate whose lattice can be changed by applying an electric field. In this way, the topological order of the semiconductor can be tuned with electric field.

\section{Summary}

In summary, we have explored the strain tuning of topological band order in cubic semiconductors by the combination of simple tight-binding analysis and density functional theory calculation. We have predicted that InSb can realize the topological insulating state under the biaxial lattice expansion of $2\%-3\%$. This work provides a generic guiding principle for tuning the topological order of the cubic semiconductors and offers an opportunity for experimentally exploring the properties of the topological surface states on such technologically relevant materials for practical applications.

\acknowledgements

This work is supported by the Laboratory Directed Research and Development Program of Oak Ridge National Laboratory, managed by UT-Battelle, LLC, for the U. S. Department of Energy.

\end{document}